\documentclass{aa}
\usepackage[varg]{txfonts}
\usepackage{newtxtext,newtxmath}

\usepackage[T1]{fontenc}
\usepackage{ae,aecompl}

\usepackage{graphicx}	
\usepackage{amsmath}	
\usepackage{amssymb}	
\usepackage[table]{xcolor}
\bibpunct{(}{)}{;}{a}{}{,} 

\newcommand{\cii}{[C\,{\sc ii}]\xspace}
\newcommand{\oiii}{[O\,{\sc iii}]\xspace}
\newcommand{\hii}{H\,{\sc ii}\xspace}

\begin{document}

\title{Highly-ionized gas in lensed $z = 6.027$ Little Red Dot seen through [O\,{\sc iii}]\,88$\mu$m with ALMA}

\author{Kirsten K. Knudsen\inst{1}\thanks{E-mail: kirsten.knudsen@chalmers.se}, 
Johan Richard\inst{2}, 
Mathilde Jauzac\inst{3,4,5,6}, 
Tom J.L.C. Bakx\inst{1}, 
Thiago S.~Gon\c calves\inst{7},
Eiichi Egami\inst{8}, 
Kiana Kade\inst{1}, 
Rahul Rana\inst{1},
Laura Sommovigo\inst{9,10}
Flora Stanley\inst{11}, \and
Daniel P.~Stark\inst{8}
}
\institute{
Department of Space, Earth and Environment, Chalmers University of Technology, SE-412\,96 Gothenburg, Sweden \and
CRAL, Observatoire de Lyon, Universit\'e Lyon 1, 9 Avenue Ch. Andr\'e, F-69561 Saint Genis Laval Cedex, France \and
Centre for Extragalactic Astronomy, Durham University, South Road, Durham DH1 3LE, U.K. \and
Institute for Computational Cosmology, Durham University, South Road, Durham DH1 3LE, U.K. \and
Astrophysics Research Centre, University of KwaZulu-Natal, Westville Campus, Durban 4041, South Africa \and
School of Mathematics, Statistics \& Computer Science, University of KwaZulu-Natal, Westville Campus, Durban 4041, South Africa \and
Observat\'orio do Valongo, Universidade Federal do Rio de Janeiro, Ladeira do Pedro Ant\^onio 43, Sa\'ude, Rio de Janeiro, RJ 20080-090, Brazil \and 
Steward Observatory, University of Arizona, 933 N Cherry Ave, Tucson, AZ 85721, USA 
\and
Department of Astronomy, Columbia University, 550 W 120th St, New York, NY 10025, USA
\and
Center for Computational Astrophysics, Flatiron Institute, 162 5th Avenue, New York, NY 10010, USA
\and 
Institut de Radioastronomie Millim\'etrique (IRAM), 300 Rue de la Piscine, F-38400 Saint-Martin-d’H\`eres, France
}

\authorrunning{Knudsen et al.} 

\date{Received ... / Accepted ...} 

\abstract{
Determining the physical properties of galaxies during the first billion
years after the big bang is key to understanding both early galaxy
evolution and how galaxies contributed to the epoch of reionization.  
We present deep ALMA observations of the redshifted \oiii 88\,$\mu$m line for
the gravitationally lensed ($\mu = 11.4\pm1.9$) galaxy A383-5.1 ($z=6.027$) that has previously
been detected in \cii 158\,$\mu$m.  
Recent \textit{James Webb Space Telescope} (\textit{JWST}) imaging identified this sub-L$^\star$ galaxy as a ''Little Red Dot'' (LRD). 
With a line luminosity of $L_{\rm [OIII]} = 
(1.29\pm0.24)\times10^8$\,L$_\odot$ (corrected for lensing magnification) A383-5.1 is one of the faintest galaxies with combined \cii{} and \oiii{} detections.
The ALMA data reveal no dust continuum emission, consistent with previous
observations.  The high line luminosity ratio of \oiii / \cii $\sim 14\pm5$ is
consistent with A383-5.1 being low-metallicity and dust-poor.  
The non-detection of dust continuum in bands 6 and 8 is consistent with the high \oiii /\cii ratio and suggests a presence of a strong ultraviolet radiation field, which would be less affect by dust attenuation, implying that galaxies of this type could contribute significantly to the ionization of the intergalactic medium. The presence of strong ionizing field could provide an important piece of information for understanding the nature of LRDs and their role in cosmic reionization.
}

\keywords{Galaxies: high-redshift -- Galaxies: evolution -- Galaxies: ISM --
Galaxies: individual: A383-5.1 -- Submillimeter: galaxies}
\maketitle 
%

\section{Introduction}

The first billion years after the big bang mark the rise in galaxy formation
and galaxy evolution, which is witnessed for example by the increased 
integrated star formation rate 
\citep[e.g., summarised in reviews, ][]{madau14,stark16}. 
The ultraviolet (UV) emission produced
in galaxies ionize the neutral hydrogen of the integalactic medium, marking
the phase transition from neutral to ionized medium at cosmological scales,
also known as the epoch of reionization (EoR). The end of the EoR is
estimated to be around redshift $z\sim 6$ \citep[e.g.,][]{stark16,planck2016}. 
It has long been debated whether the dominant sources of ionizing
UV-photons are the bright, but rarer quasars and active galactic nuclei
(AGN), or the less luminous, but more common, normal to
low-luminosity galaxies \citep[e.g.][]{fontanot12,grissom14,hassan18}.  
Observations of the luminosity
function at $z>5$ show an increased significance of the faint-end compared to
the bright-end, supporting the claim that the dominant fraction of UV-photons could
arise from the common low-luminosity galaxies
\citep[e.g.][]{mclure13,bouwens16,parsa18}. 

Determining the physical properties of galaxies during and at the end of the
EoR provides necessary constraints for understanding the nature of the sources
that could be responsible for the bulk of the UV-emission giving rise to the
re-ionisation. This also provides insights for
characterising the early galaxy evolution.  With the increasing number of
galaxies with spectroscopic redshift at $z>6$
\citep[e.g.][]{schenker12,shibuya12,finkelstein13,oesch15,zitrin15}, 
it is possible to study the physical properties using, for example, the
far-infrared fine-structure lines \cii and \oiii
\citep[e.g.][]{ouchi13,ota14,maiolino15,knudsen16,inoue16,bradac17,laporte17,hashimoto18,fujimoto21,jolly21,bouwens22,fudamoto22,algera23,fujimoto24}
and search for dust emission
\citep[e.g.][]{watson15,laporte17,laporte21,sommovigo22b,inami22}. 

The ionization potential of neutral oxygen is similar to that of hydrogen,
and the ionization potential of ionized oxygen is 35\,eV.  Double ionization of
oxygen requires a UV-radiation field characteristic of O-type stars and
that of AGN.  As a result, doubly ionized oxygen is typically found in \hii
regions. Double ionized oxygen has two far-infrared fine-structure lines at
52\,$\mu$m and 88\,$\mu$m. The latter is observed to be very bright in
star-forming galaxies, at a line intensity similar to that of the \cii
158\,$\mu$m line \citep[e.g.][]{delooze14}, and in the low-metallicity dwarf galaxies
it is seen to be brighter than the \cii line \citep[e.g.][]{cormier15}.  
This makes it a very attractive line to study in $z>6$ galaxies, as it is
also redshifted to submm wavelengths, and  \oiii 88\,$\mu$m is
detected  in a small number $z>7$ galaxies 
\citep[e.g.][]{inoue16,laporte17,hashimoto18,harikane20}.

In this paper we present Atacama Large Millimetre/Submillimetre Array (ALMA)
observations of \oiii\,88$\mu$m in the sub-$L^\ast$, gravitationally lensed
galaxy A383-5.1 ($z = 6.027$). 
With the combination of the extensive ALMA integration time invested to this
project and the gravitational lensing magnification, which is estimated to
be $\mu = 11.4\pm1.9$ \citep{richard11,stark15}, this is among the deepest
observations for a single high-$z$ source carried out with ALMA. 
Based on an analysis of the optical and near-infrared spectroscopy and
photometry of the counter-image A383-5.2, the system has an estimated SFR of
$2.0^{+0.34}_{-0.3}$\,M$_\odot$\,yr$^{-1}$, a stellar mass of $M_{\rm
stellar} = (3.2^{+0.8}_{-0.7})\times10^9$\,M$_\odot$, and a very bright Ly$\alpha$
line, which could imply a somewhat higher SFR \citep{richard11,stark15}. 
Furthermore, the modelling from \citet{stark15}
shows that the galaxy is best described by a two-component star formation
history, where a young starburst component contributes only little to the
stellar mass, but contributes significantly to the ionizing radiation. 
Also, the counter-image A383-5.2 has a near-infrared spectroscopic detection of
C\,{\sc iii}], and A383-5.1 has one of the lowest-luminosity \cii detections
published so-far \citep{richard11,stark15,knudsen16}.

New {\it James Webb Space Telescope (JWST)}/NIRCam observations reveal that rest-frame optical images of A383-5.1 and 5.2 consists of two components, namely a 'Little Red Dot' (LRD) and a blue component \citep{golubchik2025,baggen2025}. LRDs are an extremely red population of sources found in \textit{JWST} observations \citep[e.g.,][]{Kocevski2023, Labbe2023, labbe2025}. These galaxies exhibit unusual properties including distinctive V-shaped spectral slopes, broad emission lines, generally lacking X-ray emission, and no dust and radio detections \citep[e.g.,][]{Kocevski2023, Labbe2023, labbe2025, Greene2024, Casey2025, Maiolino25, Perger2025}. Providing a clear explanation for these properties has been challenging. A leading explanation is the 'Black Hole *' (BH*) model wherein the AGN at the center of the LRD is surrounded by a dense screen of hydrogen \citep{Naidu2025}. The SED of one component of the A383-5.1 system exhibits the V-shape characteristic of LRDs \citep{golubchik2025,baggen2025}, although it should be noted that the LRD classification is tentative until spectroscopic confirmation can be carried out. \citet{golubchik2025} found that the SED of the LRD component was best fit by a BH* model whereas the blue component was found to be most consistent with a young star-forming nebular object. \citet{golubchik2025} suggested that a possible explanation of this system is that the blue object is metal-enriched emission that was ejected from the red component hosting the LRD. \citet{baggen2025} suggested that the blue component is a young star-forming population, while the red component could be fit with a massive and dusty source. However, they also discuss that the red component could be interpreted as a combination of stellar and AGN emission. The detection of \cii and \oiii in A383-5.1 are the first detections of far-infrared emission lines in an LRD.
The ALMA data presented here provide additional input to the discussion on the nature of A383-5.1. 

We assume a $\Lambda$CDM cosmology with $H_0 =
67.4$\,km\,s$^{-1}$\,Mpc$^{-1}$, $\Omega_M = 0.315$, and $\Omega_\Lambda =
0.685$ \citep{planck20}.  

%

\section{Observations}

We have obtained ALMA observations of the redshifted \oiii 3393.00624\,GHz
line in A383-5.1 (project ID 2016.1.00333.S). 
The observations were carried out using the band-8
receiver during cycle-4 and 5 \citep{Sekimoto2008}.  Ten executions were done resulting in a total
on-source integration time of 8 hours.  
For the receivers one spectral window was tuned to redshifted \oiii line
based on the redshift derived from the \cii detection \citep{knudsen16} with
a bandwidth of 1.875\,GHz and spectral resolution of 3.906\,MHz (using
Frequency Division Mode).  The three 
other available spectral windows used a continuum setup with a bandwidth of
2\,GHz each distributed over 128 channels (using Time Division Mode).
The telescope configuration has baselines extending between 15 and 783\,m. 

The Common Astronomy Software Application \citep[CASA][]{mcmullin07} was used
for reduction, calibration, and imaging. 
The pipeline reduced data delivered by the observatory was of sufficient
quality, therefore no additional flagging and further calibration were done. 
The ALMA pipeline includes the steps required for standard reduction and
calibration, such as flagging, bandpass calibration, as well
as flux and gain calibrations.  
A conservative estimate of the absolute flux calibration is $10\%$, based on the ALMA Technical Handbook \citep{ALMATECHNICALHANDBOOK2019}. 

Imaging both the continuum and spectral cube using natural
weighting the resulting resolution is $0.37'' \times 0.32''$ PA\,$=-84^\circ$
and the r.m.s.\ is 35\,$\mu$Jy\,beam$^{-1}$ and 0.41\,mJy\,beam$^{-1}$ (in a
$\sim20$\,km\,s$^{-1}$ channel), respectively. 

Moreover, we extract additional \cii data from the ALMA archive from
project 2015.1.01136.S, which had 60\% of the observing time compared to our
old \cii data, but does add twice as long baselines.  Data were reduced
following standard procedures, and then combined with our previous data, resulting
in an angular resolution of $0.33''\times 0.29''$ PA\,$=-80^\circ$ and an rms
of 0.17\,mJy\,beam$^{-1}$ (in a $\sim17$\,km\,s$^{-1}$ channel).

%

\section{Results}

\begin{figure}
\centerline{\includegraphics[width=7.0cm]{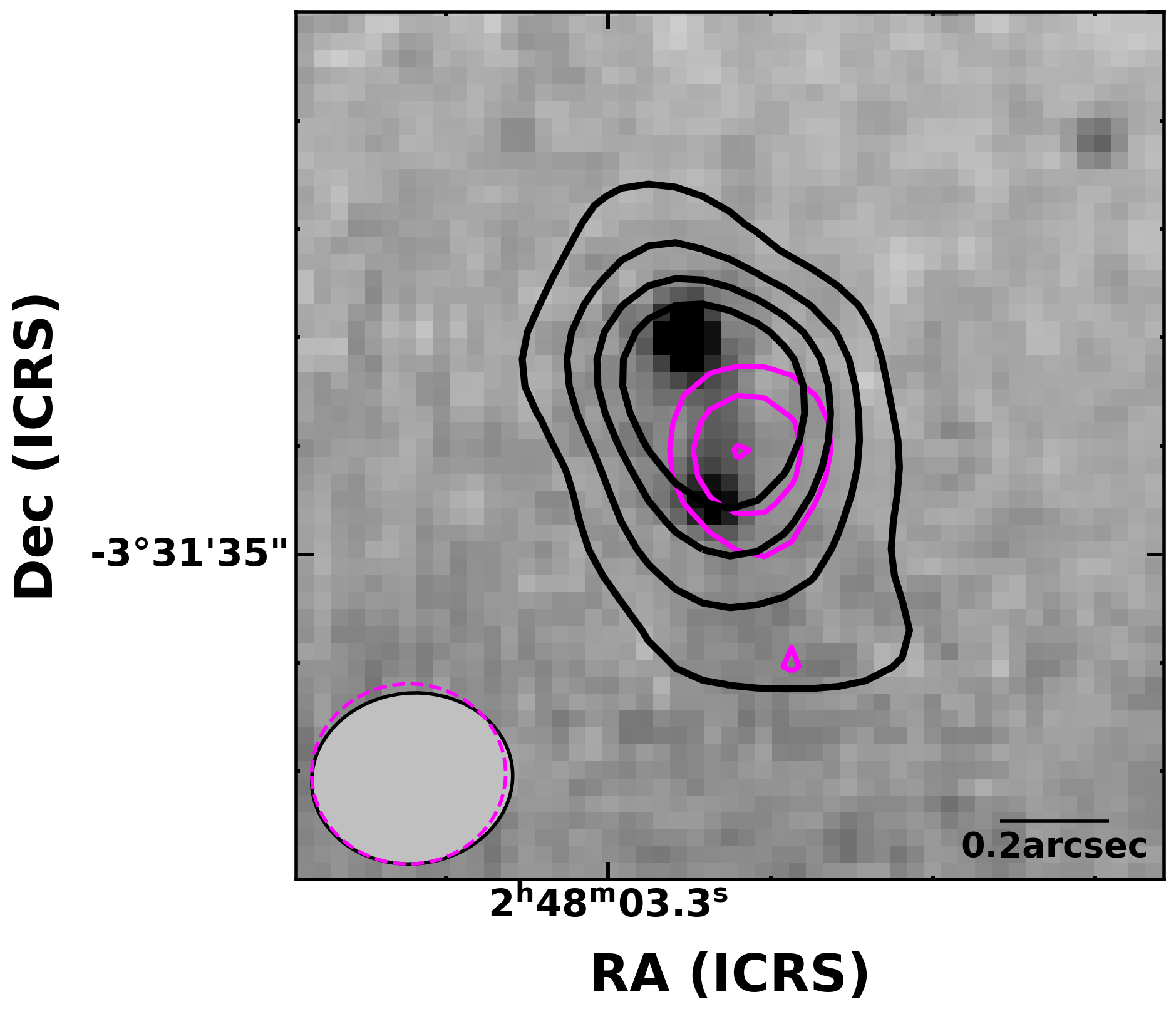}}
\centerline{\includegraphics[width=8.0cm]{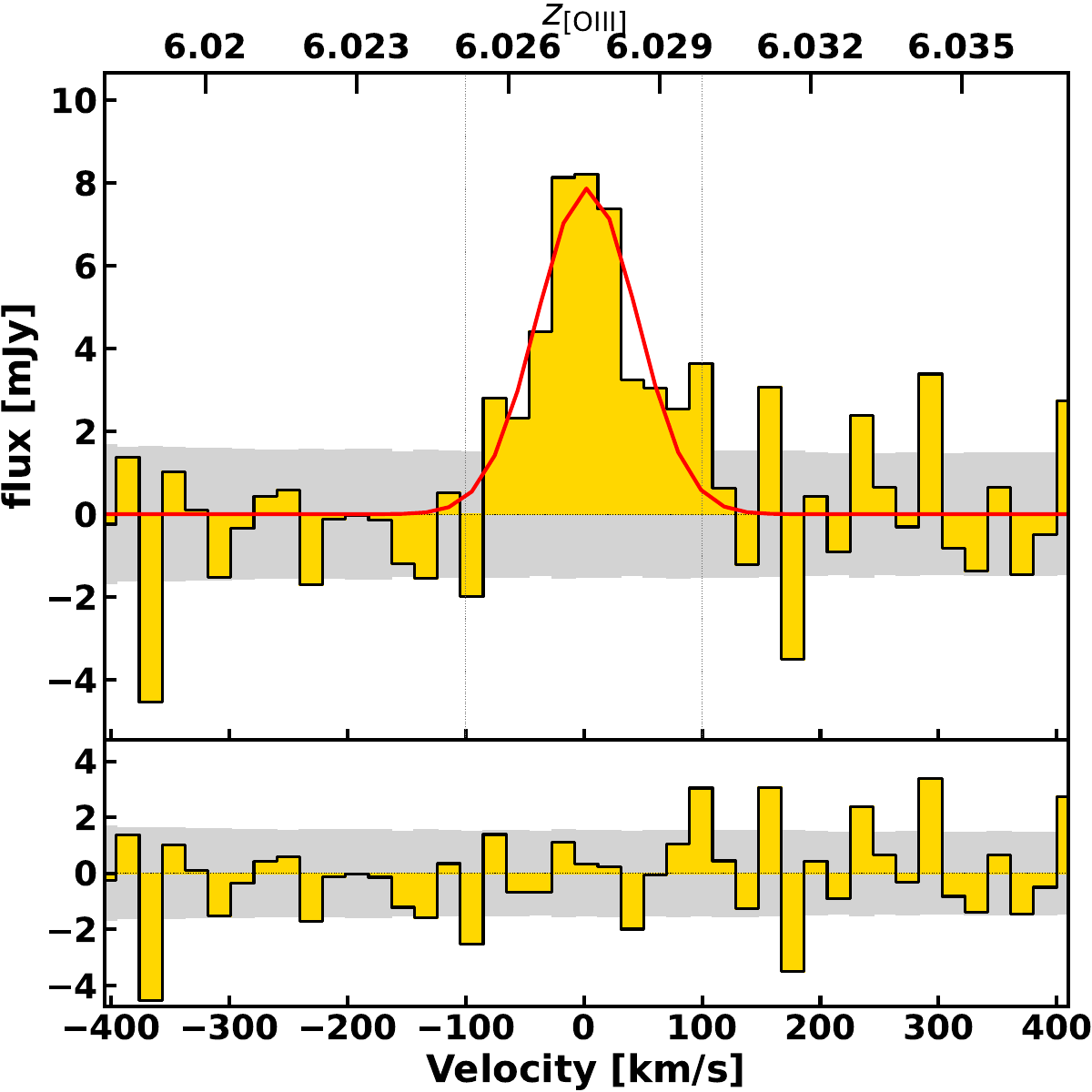}}
\caption[]{The \oiii 88\,$\mu$m detection of A383-5.1.  
{\it Top}: Moment-0 map obtained through collapsing the data cube over
the velocity range -100 to 100\,km/s. 
The contours show $3, 5, 7, 9\sigma$, and dashed
show $-3\sigma$ overlaid on the \textit{JWST} NIRCam/F200W image. Additionally, the magenta contours show the moment-0 of \cii using the combined data from \citet{knudsen16} and additional ALMA archive data;  \cii contours show $3, 4, 5\sigma$. 
{\it Bottom}:  
ALMA spectra extracted at the position of A383-5.1 and centered at the
frequency of the redshifted \oiii line.  The red dashed line shows the
best-fit Gaussian.  
The vertical lines indicate the velocity range over which the moment-0 map
has been extracted. 
The lower panel shows the residuals after subtracting the best-fit Gaussian profile from the spectrum.  
\label{fig:a383result} }
\end{figure}

We detect the redshifted \oiii line in A383-5.1.
We extract the spectrum based on the $3\sigma$ level of the moment-0 map,
and the peak is detected at 12$\sigma$.  
The spectrum and the moment-0 map are shown in Figure~\ref{fig:a383result} (the moment-0 contours are overlaid on the NIRCam F200W image), and the
resulting parameters of a single Gaussian line profile are 
$S_{\rm peak} = 7.9\pm1.1$\,mJy, linewidth FWHM = $99\pm15$\,km\,s$^{-1}$,
which corresponds to an integrated line intensity of
$0.83\pm0.16$\,Jy\,km\,s$^{-1}$.  The estimated redshift is
$z=6.0275\pm0.0002$, consistent with the \cii redshift of
$z=6.0274\pm0.0002$.
Similarly, the linewidth as given by FWHM is consistent
with that derived for the \cii line of $100\pm23$\,km\,s$^{-1}$
\citep{knudsen16}. Results are summarized in Table~\ref{tab:results}.

The estimated size given as the FWHM of the spatial distribution, based on a
two-dimensional fit to the moment-0 map, is $(0.58\pm0.07)'' \times (0.22\pm
0.07)''$ (PA = 20$^\circ$), corresponding to an area of 0.1\,arcsec$^2$. 
Correcting for gravitational lensing magnification, this corresponds to an
area of 0.3\,kpc$^{2}$.  
This is consistent with the {\it HST} WFC3/F110W continuum
imaging results from \citet{richard11}, who noted that the Lyman-$\alpha$
emission is more extended. 

In \citet{knudsen16}, 
the angular resolution of the \cii detection
was not sufficient to resolve the line emission. Combining with the archival
data with similar depth but longer baselines\footnote{Project number:
2015.1.01136.S}, we find a
similar result for the line profile.  We note that the 
better {\it uv}-coverage shows also extended \cii. 
A two-dimensional Gaussian fit yields $(0.38\pm0.09)'' \times
(0.24\pm 0.09)''$ (PA = 8.8$^\circ$), consistent with the \oiii result.  
We note that the \cii detection has a lower significance than the
\oiii one, and therefore the accuracy of the size of the \cii emission
is lower. 

\begin{table}
\centering
\caption[]{Observed and estimated properties from the \oiii line of A383-5.1}
\begin{tabular}{lc}
\hline
\hline
Parameter &  Value \\
\hline 
$z_{\rm [OIII]}$ & $6.0275\pm0.0002$ \\
$S_{\rm peak}$ & $7.9\pm1.1$\,mJy \\
FWHM &  $99\pm15$\,km\,s$^{-1}$ \\
$I_{\rm [OIII]}$ & $0.83\pm0.16$\,Jy\,km\,s$^{-1}$ \\
$L_{\rm [OIII]}$ & $(1.29\pm0.24)\times10^8$\,L$_\odot$ \\
SFR$_{\rm [OIII]}$ & 5.6\,M$_\odot$\,yr$^{-1}$ \\
\hline
\end{tabular}
\tablefoot{ 
   The line luminosity and SFR are corrected for
gravitational lensing magnification.} 
\label{tab:results}
\end{table}

We estimate the \oiii line luminosity using $L_{\rm line} = 1.04\times 10^{-3} S
\Delta V D_{\rm L}^2 \nu_{\rm obs}$
\citep[e.g.][]{solomon05,carilliwalter13} to be $L_{\rm [OIII]} =
(1.29\pm0.24)\times10^8$\,L$_\odot$ corrected for lensing magnification; the
uncertainty reflects the propagated error from the Gaussian fit to the line
profile.  
\citet{delooze14} have investigated the \oiii line as a SFR estimator for local
star-forming galaxies against other probes, and for different classes of
galaxies.  Using the relation for low-metallicity galaxies we estimate
SFR$_{\rm [OIII]} \sim 5.6$\,M$_\odot$\,yr$^{-1}$, which is above other
previous estimates for the SFR based on optical and \cii results
\citep{richard11,stark15,knudsen16}. 
The estimated \cii line luminoity is $L_{\rm [CII]} = (8.9\pm3.1)\times10^6$\,L$_\odot$ 
\citep{knudsen16}, which implies a line luminosity ratio of
\oiii\!/\,\cii$\sim14\pm 5$.

Continuum emission is undetected in the band-8, and we place a $3\sigma$
upper limit of $0.105$\,mJy. 
We use this together with the upper limit from the band-6 observations to 
estimate an upper limit of the far-infrared luminosity.  Assuming a modified
black-body spectral energy distribution with a temperature $T = 35$\,K and
$\beta = 1.6$ 
(and correcting for the CMB radiation field), we find $L_{\rm
FIR} < 5.6\times10^{9}$\,L$_\odot$.  This is consistent with the upper limit
based on the band-6 observations from \citet{knudsen16}. For higher $T$, the
upper limit on $L_{\rm FIR}$ will approach $<10^{10}$\,L$_\odot$. 
Similarly, if selecting a slightly higher $\beta$ will similarly increase the
upper limit; results from \citet{faisst20} suggests $\beta$ values between
1.6 and 2.4 for high-$z$ galaxies. 
Assuming the total IR luminosity to be about 30\% higher
\citep[e.g.][]{decarli17}, and 
assuming a conversion ${\rm SFR}[{\rm M}_\odot\,{\rm yr}^{-1}] =
1.07\times10^{-10}\,L_{\rm IR} [{\rm L}_\odot]$ \citep[assuming a Kroupa IMF
in the mass range 0.1-100\,M$_\odot$ and a timescale of constant star
formation of $\tau=100$\,Myr]{calzetti13}, this
would correspond to an SFR\,$<0.4$\,M$_\odot$\,yr$^{-1}$ (or
$<0.8$\,M$_\odot$\,yr$^{-1}$ if using $T=45$\,K), which is below that
of optical and \oiii-based estimates; keeping in mind a higher dust temperature would imply a
higher upper limit on the $L_{\rm FIR}$ and thus also a higher upper limit on
the SFR.  This is implies that the obscured fraction of star formation is small. 

\begin{figure}
\centerline{\includegraphics[width=0.9\columnwidth]{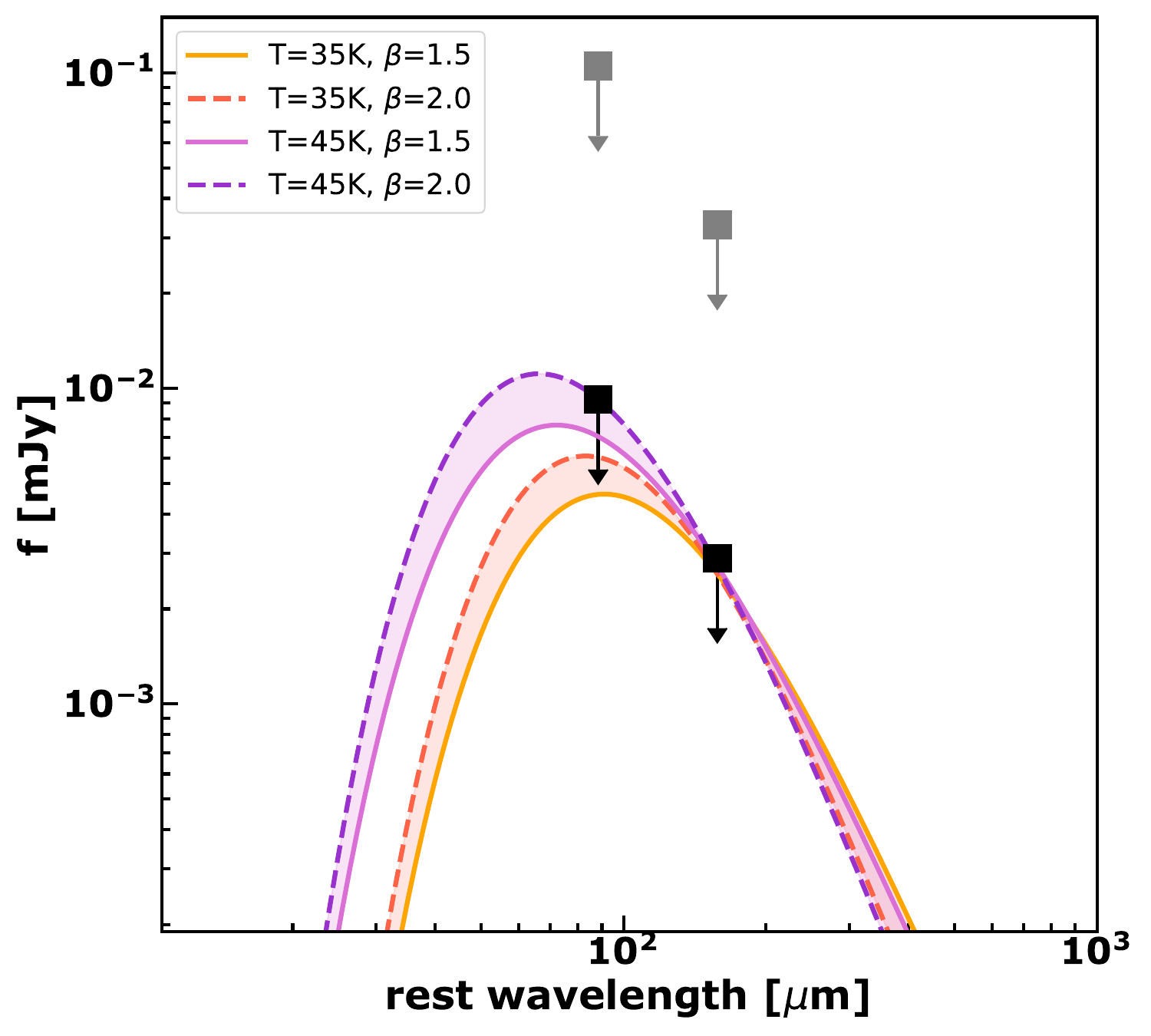}}
\caption{
The far-IR SED of A383-5.1 with the two continuum upper limits from bands 6 and 8.  The grey squares show the $3\sigma$ observed limits, while the black squares show the same corrected for gravitational lensing magnification. The modified blackbody functions are normalised to the band-6 data point.  
\\
\label{fig:FIRSED}
}
\centerline{\includegraphics[width=0.75\columnwidth]{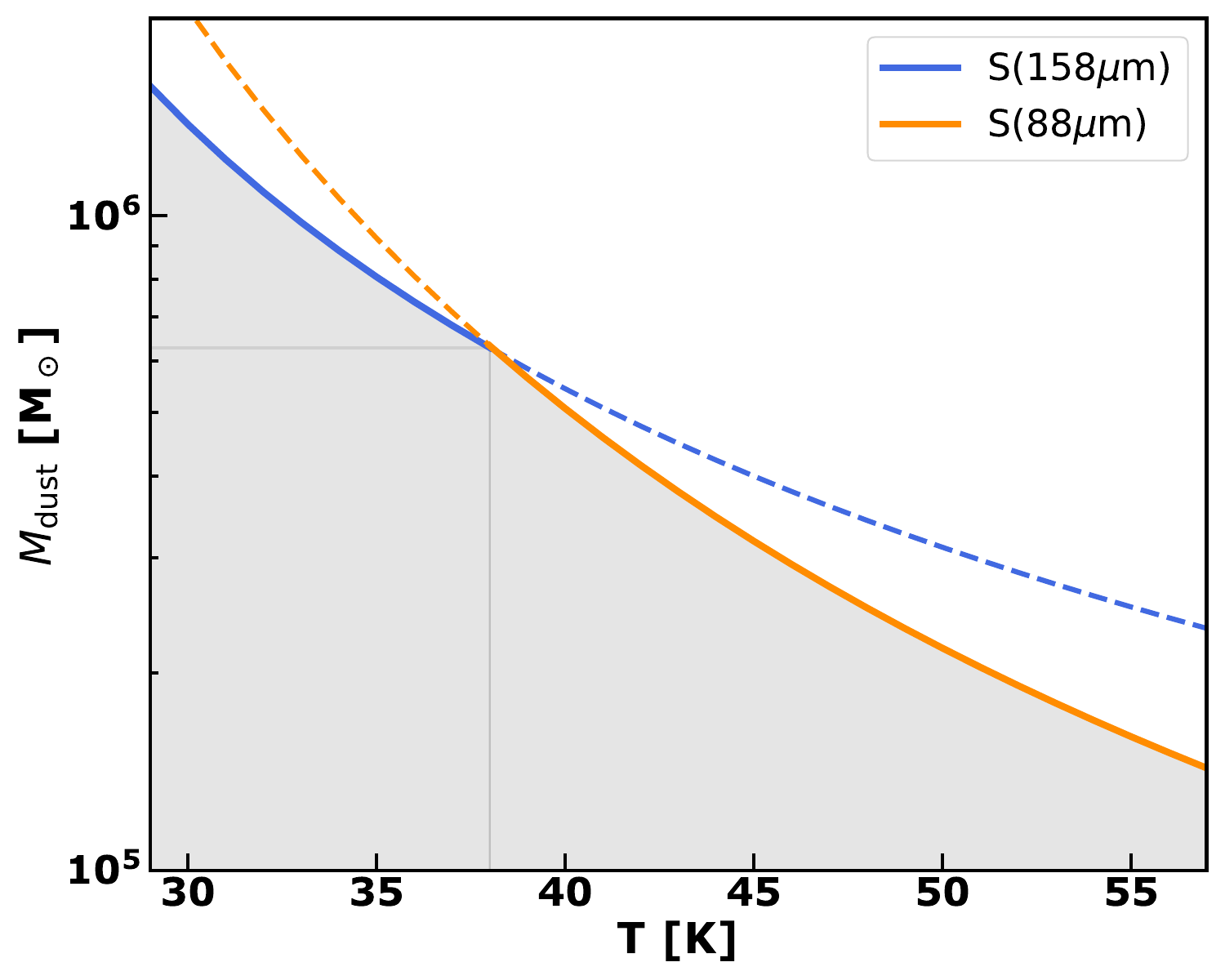}}
\caption{
Estimate on the $3 \sigma$ upper limit of the dust mass as a function of temperature $T$ and with the assumed $k_\nu$ values given in the text. The blue [orange] line shows the limit when normalising to the band-6 [band-8] upper limits.  When normalising to one of the upper limits, it places a constraint on the temperature, which is indicated with the dashed line.  The grey area indicates the range of dust masses as a function of the dust temperature between $T = 30 - 55$\,K. Along this range, the mass limit changes by more than a factor 8. 
\label{fig:mdust}
}
\end{figure}

With two continuum upper limits in band-6 and band-8 it is not possible to
constrain the far-infrared SED shape. However, we can
estimate an upper limit on the dust mass.  We use the $3\sigma$ limit
from the band-6 data of $f_{\rm \nu_{\rm obs} = 270\,GHz} < 0.033$\,mJy and
the band-8 upper limit, and
assume that the dust follows local galaxy properties with an absorption
coefficient of $\kappa_{\rm 88\mu m} = 44.1$\,cm$^2$\,g$^{-1}$ and
$\kappa_{\rm 158\mu m} = 12.0$\,cm$^2$\,g$^{-1}$
\citep[e.g. ][]{draine07,draine14}.  
For $S_{\nu_{\rm obs}} \propto \kappa_{\nu_{\rm rest}} B_{\nu_{\rm rest}}(T) M_{\rm d}$, where
$\nu_{\rm rest}$ is the rest-frame frequency and $\nu_{\rm obs}$ is the redshifted frequency, we assume $T=35$\,K
and estimate an upper limit on the dust mass $M_{\rm d} <
0.8\times10^6$\,M$_\odot$ (corrected for lensing magnification and the effect of the CMB);  
a higher dust temperature would result in an even lower dust mass limit. Figs.~\ref{fig:FIRSED} and \ref{fig:mdust} show the FIR upper limit on the continuum together with modified blackbody functions for a range of temperature and $\beta$, and the upper limit on the dust mass as function of temperature using the two upper limits.  While each individual upper limit does not provide a solid upper limit, together they provide a marginally improved constraint if fixing the modified blackbody to one of the photometric upper limits. 
The dust mass upper limit is similar to the dust mass upper limits for sample 
of stacked gravitationally lensed $z\sim 6$ galaxies \citep{jolly21}.
Without knowledge of dust properties and dust temperature, this estimate
remains uncertain, nonetheless it shows that A383-5.1 is a dust-poor system.
This deep upper limit on the dust mass results in valuable constraints on the nature of A383-5.1. If the rest-frame optical emission of this source is from a star-forming galaxy, the galaxy mass is estimated around $\log_{10} M_\star/{\rm M_\odot} = 8.9$, while the presence of an optically-luminous AGN would reduce this stellar mass to $\log_{10} M_\star/{\rm M_\odot} = 7.7$ \citep{golubchik2025}. Assuming a dust temperature of around 40\,K, the subsequent dust-to-stellar mass ratios in these two scenarios vary between an extremely dust-poor ($< 3 \times10^{-4}$) galaxy, and a low dust presence of ($< 7\times10^{-3}$) in the interstellar medium surrounding an AGN system.

%


\section{Discussion}

The very high magnification factor of A383-5.1 has allowed for the first ALMA detection of far-infrared fine structure emission lines to be detected in a low stellar mass galaxy hosting an LRD\footnote{\citet{Xiao2025} did targetted ALMA observations of two massive and bright LRDs and placed upper limits on the \cii line luminosity and thermal dust luminosity.}. 
The observed high line ratio combined with the LRD identification suggests that this galaxy is not akin to local analogs (e.g., low-metallicity dwarf galaxies, e.g., \citealt{cormier15}), but is rather part of the new extreme population revealed by \textit{JWST}. Therefore, we begin by discussing the high line ratio in this context.

\begin{figure}
\centerline{\includegraphics[width=0.99\columnwidth]{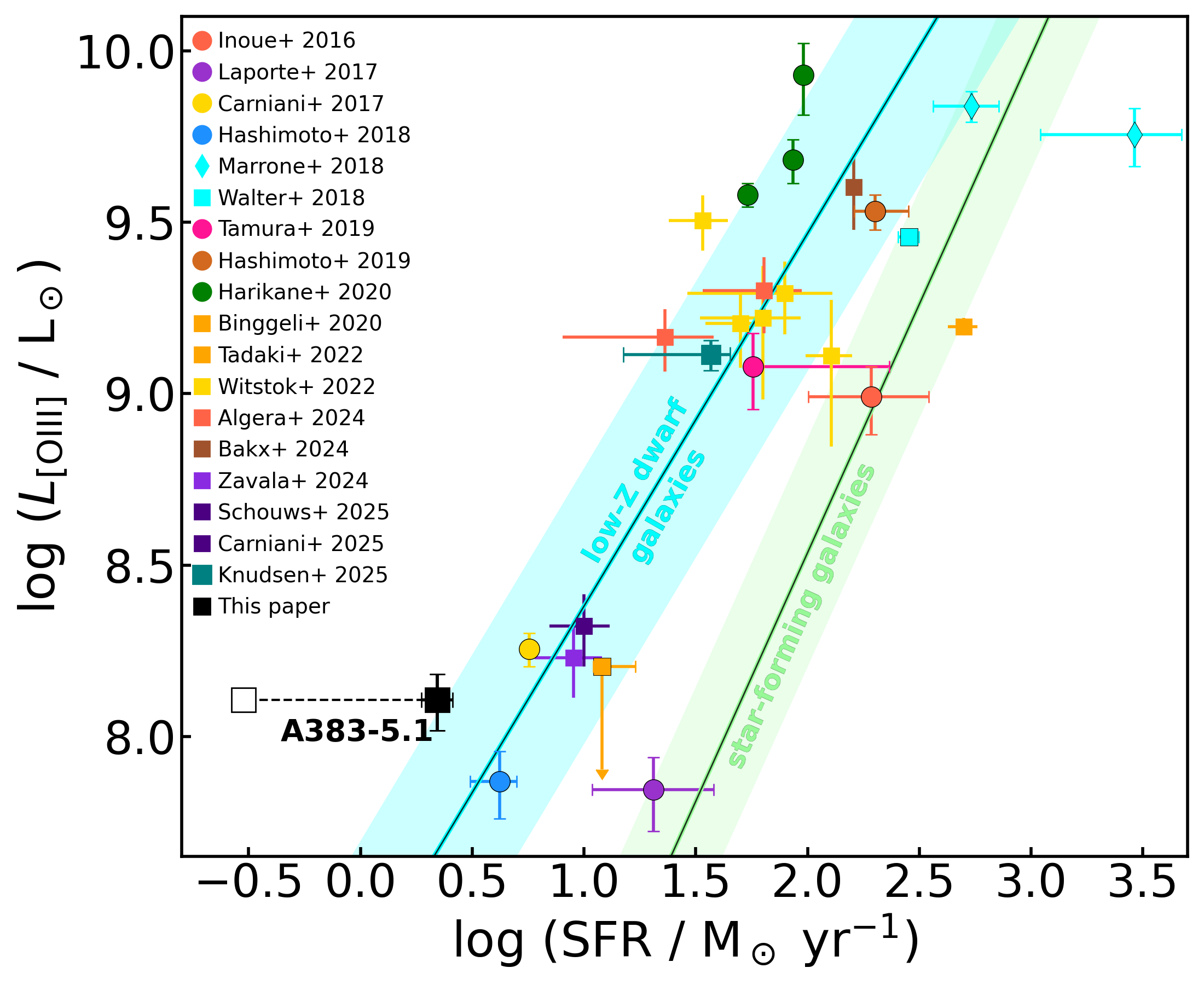}}
\caption[]{
\oiii line luminosity, $L_{\rm [OIII]}$, vs.\ star formation rate. 
The filled black square shows the detection in A383-5.1 using the SFR of \citet{richard11}, and the open black square shows the same but with the SFR limit of \citet{golubchik2025}, while the coloured symbols show 
recent results 
\citep{inoue16,carniani17,laporte17,marrone18,hashimoto18,tamura19,hashimoto19,harikane20,binggeli20,akins22,algera23,tadaki22,witstok22,bakx2024,zavala24,schouws25,carniani25,knudsen2025}.
The $L_{\rm [OIII]}$ - SFR relation, where 
The green region shows the $L_{\rm [OIII]}$ - SFR relation
for local star-forming galaxies, and the light-blue one shows that for
low-metallicity dwarf galaxies \citep{delooze14}.
\label{fig:LoiiiSFR}}
\end{figure}

The \oiii detection, and the previous \cii detection, are the first far-infrared fine-structure line detections in an LRD.  Focusing  first on how the line properties relate to star formation measurements, we compare to local galaxies, where a relation is seen between SFR and far-infrared
fine-structures lines \cii 158$\mu$m, \oiii 88$\mu$m, and [O\,{\sc i}]\,63$\mu$m
\citep[e.g.][]{delooze14}.  
In the case of the \oiii line, there appears to be
up to an order of magnitude difference in the estimated SFR-line luminosity
relation depending on properties such as the metallicity. 
Based on an analysis prior to {\it JWST} observations, the second lensed image, A383-5.2, has an estimated metallicity of $Z\sim(0.043^{+0.030}_{-0.013}) Z_\odot$ and mass of
$\sim3\times10^9$\,M$_\odot$ \citep{richard11,stark15}, indicating that the system could be similar to local low-$Z$ dwarf galaxies.
Low-metallicity
dwarf galaxies have a significantly 
higher average \oiii line luminosity than starbursts for similar SFR. 
Assuming the estimated SFR of
$2.0^{+0.34}_{-0.3}$\,M$_\odot$\,yr$^{-1}$ \citep{richard11,stark15}, A383-5.1 follows this trend within the scatter of the local relation for low-$Z$ dwarf galaxies. 
In Fig.~\ref{fig:LoiiiSFR}, we plot the $L_{\rm [OIII]}$ vs.\ SFR together
with other $z>6$ \oiii detections. 
But contrasting this with the discussion of the SFR based on the new {\it JWST}/NIRCam results, which suggests that the SFR cannot exceed 0.5\,M$_\odot$\,yr$^{-1}$ \citep{golubchik2025}, this would indicate that the \oiii is to a large extend not powered by star formation, but rather by some other mechanism (e.g., by ionization from the radiation field from a growing super-massive black hole); the location of the NIRCam-estimated SFR is also indicated in Fig.~\ref{fig:LoiiiSFR} and shows how the \oiii line luminosity exceeds the local relation between SFR and $L_{\rm [OIII]}$.

The specific SFR, ${\rm sSFR} = {\rm SFR}/M_{\rm stellar}$, is
$\sim 10^{-8}$\,yr$^{-1}$ assuming the values including the BH* interpretation of the LRD \citet{golubchik2025}; the sSFR would be lower, though assuming a model without BH*.  As discussed in \citet{algera2024,algera2025b}, a high \oiii /\cii ratio could be indicative of a system of high burstiness. The sSFR together with the line ratio could suggest that components of A383-5.1 is a young system, which would be in agreement with the SED modelling results for the blue companion \citep{golubchik2025,baggen2025}. We note the large uncertainties in the sSFR, though, making it challenging to make conclusive statements on this.  

The new {\it JWST}/NIRCam imaging shows that A383-5.1 resolves into two sources, a red compact source hosting an LRD, and a blue companion source \citep{golubchik2025}, with a possible indication of bridging emission between the components \citep{baggen2025}. 
The distribution of the \oiii emission relative to the \textit{JWST}/NIRCam imaging is shown in Fig.~\ref{fig:a383result}. Generally the distribution of the \oiii emission follows that of the overall structure of the entire system from the \textit{JWST} imaging, although we note the angular resolution of the ALMA observations is lower than that of the NIRCam F200W image. The \cii observations, which have a lower signal-to-noise ratio, are similarly limited by the angular resolution, but appear to be centered between the blue and red component \citep{golubchik2025}. Due to the limited angular resolution and signal-to-noise ratio, we cannot currently constrain if only a single component in the system is responsible for the majority of one or both of the far-infrared emission lines, and therefore not determine if the emission is powered by star formation or a rapidly growing supermassive black hole.  

The population of LRDs has been found to be dust poor \citep{Casey2025}, which is also the case for A383-5.1 with no dust emission detection in both ALMA band 6 and 8 observations (corresponding to the \cii and \oiii measurements). The observed low dust content in A383-5.1 implies a low dust attenuation of the UV radiation. This is consistent with the bright \oiii line observations, as photons with energy $>$\,35\,eV are required to double ionize oxygen. Generally, the possible impact of dust obscuration seem negligible, which is consistent with the estimated low metallicity in the system \citep{richard11,stark15,golubchik2025}.

\begin{figure}
\centerline{\includegraphics[width=0.90\columnwidth]{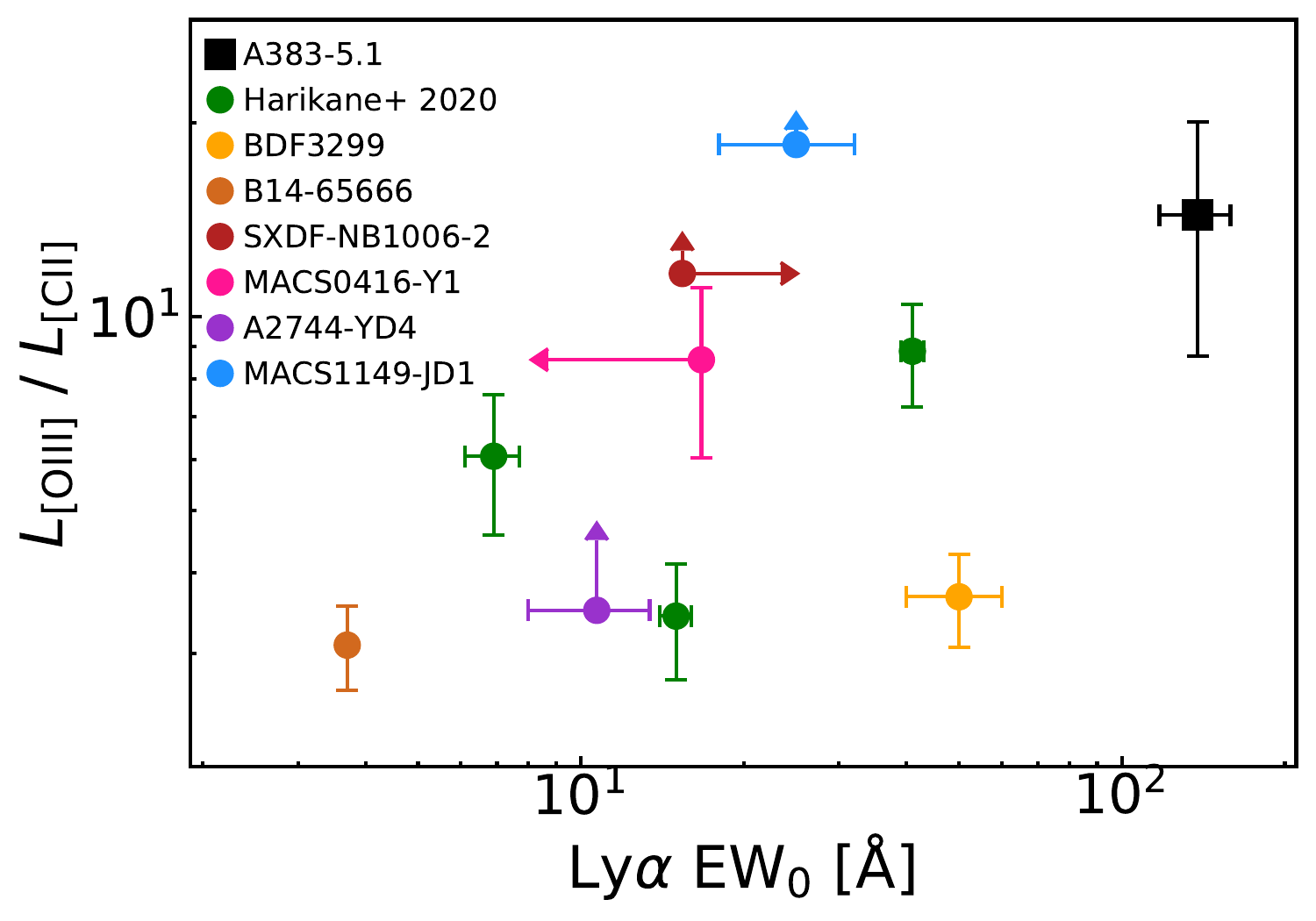}}
\caption[]{
The  \oiii\!/\,\cii line luminosity ratio vs.\ the Ly$\alpha$
equivalent width. 
Black square shows the detection in A383-5.1, and coloured symbols show 
results for other $z>6$ galaxies 
\citep{vanzella11,shibuya12,furusawa16,inoue16,carniani17,laporte17,laporte19,hashimoto18,tamura19,hashimoto19,harikane20}. 
\label{fig:lya}}
\end{figure}

With a redshift of $z=6.027$, A383-5.1 is near the end of the EoR
\citep[e.g.,][]{planck2016}, and given the large Ly$\alpha$ equivalent
width, it is likely not located in a neutral region of the intergalactic medium (IGM). 
As pointed out by, e.g., \citet{harikane20}, there appears to be a trend of
increasing \oiii /\cii line luminosity ratio with increasing Ly$\alpha$
equivalent width (EW) as seen for $z>6$ Lyman Break Galaxies. 
This trend can be interpreted as these high-$z$ galaxies having high
ionization parameters and/or low PDR covering factors, meaning that there is
a deficiency of neutral gas surrounding H\,{\sc ii} regions \citep{harikane20}. 
With the high line luminosity ratio and large
Ly$\alpha$ EW\,$\sim140$\,\AA\ \citep{stark15} A383-5.1 follows this trend; 
in Fig.~\ref{fig:lya}, we plot the \oiii /\cii line luminosity ratio as
function of Ly$\alpha$ EW for comparison with other results. 
This suggests that the neutral
gas content of sub-L$^\star$ galaxies like A383-5.1 may be even lower, which
would lead to a higher escape fraction of ionizing UV photons. If correct,
this would naturally lend support to the suggestion that sub-L$^\star$ star-forming
galaxies make a significant (and likely a dominant) contribution to the EoR
\citep[e.g., ][]{mclure13,bouwens16,parsa18,ishigaki18}. 
Even though A383-5.1 is observed near the end of the EoR, 
the properties of A383-5.1 could be similar to those of sub-L$^\star$
galaxies 100-200 million years earlier, thus providing insights to the EoR sub-L$^\star$ population. Given its low metallicity, dust content, and high degree of
ionization, the Lyman continuum escape fraction is likely significant. 

Gravitational lensing is key to explore populations of fainter galaxies, with lower stellar masses and star-formation rates. Our source, as well as the $z = 9.1$ MACS1149-JD1 \citep{hashimoto18}, allow us to characterize the sub-L$^\star$ population, best representative of the evolution of the bulk of galaxies across cosmic time. Even with magnification factors of $\mu > 9$, both sources required $\sim 10$~hr on-source time for detection with ALMA. In order to expand samples of these more lower-luminosity galaxies during the EoR, extensive observing time needs to be involved. ALMA, and its future updates including the Wideband Sensitivity Upgrade \citep{Carpenter2023}, will play an immense role in developing our understanding of the nature of the numerous, but lower luminosity galaxies in the EoR.

\section{Summary}

We report the detection of the \oiii 88$\mu$m emission line from the $z=6.027$ star-forming
galaxy A383-5.1, which based on recent {\it JWST} observations has been identified to include an LRD.  The combination of an 8-hours on-source integration with a 
gravitational lensing magnification of $\mu=11.4\pm1.9$, makes these data among the
deepest band-8 images of a $z\sim6$ source so far. 
\oiii is detected at $12\sigma$ with an integrated line intensity of
$0.83\pm0.16$\,Jy\,km\,s$^{-1}$ (the error reflects the uncertainty from the
Gaussian fit to the line profile) and a line luminosity of $L_{\rm [OIII]} = 
(1.29\pm0.24)\times10^8$\,L$_\odot$.  The redshift and line width are consistent
with that of our previous \cii detection.  The emission is resolved over an
area of 0.3\,kpc$^2$ (corrected for lensing magnification). 
The continuum non-detection is consistent with the previous band-6
non-detection, and we estimate an upper limit on the dust mass of
$0.8\times10^6$\,M$_\odot$, assuming dust properties similar to Galactic
dust, and note that for other assumptions, e.g., higher dust temperature, the
upper limit would be even lower and close to $10^5$\,M$_\odot$. 
Compared to its stellar mass, this is very low, and thus indicates that the galaxy is
extremely dust-poor.  The derived \oiii to \cii luminosity ratio is $\sim14$, 
which is large compared to normal star-forming galaxies, but consistent with
low-metallicity, low-stellar mass galaxies in the local universe.

The recent {\it JWST}/NIRCam discovery of this system containing two components, an LRD and a blue companion, would imply that A383-5.1 is among the few LRDs with successful far-infrared emission line detections. 
The high \oiii /\cii line ratio, a possible sign of a strong UV radiation field, could be provide a means for finding embedded growing black holes, if this is indeed the nature of LRDs.
The relatively bright \oiii detection together with the high \oiii /\cii line ratio also could imply that the system is experiencing a burst-like episode, especially considering the fact that there are two components and possibly a bridge of emission between the two. 

A383-5.1 is a sub-L* galaxy, and therefore in the luminosity range of
galaxies that have been proposed to contribute significantly to the
ultraviolet emission that reionizes the IGM during the epoch of reionization. 
With the remarkably low dust-mass and high \oiii /\cii line ratio, dust attenuation will not significantly
impact the escape fraction of ionizing UV-photons.  
Our results demonstrate the need for high-frequency ALMA observations for detailed studies of the lower luminosity galaxy population that likely could be responsible for a significant fraction of the reionization.

\begin{acknowledgements}
We acknowledges support from the Nordic ALMA Regional Centre (ARC) node based at Onsala Space Observatory. The Nordic ARC node is funded through Swedish Research Council grant No 2017-00648.
KKn and KKa acknowledge support from the Knut and Alice Wallenberg Foundation (KAW 2017.0292). KKn and RR acknowledge support from the ERC synergy grant 101166930 (RECAP). 
MJ is supported by the United Kingdom Research and Innovation (UKRI) Future
Leaders Fellowship 'Using Cosmic Beasts to uncover the Nature of Dark Matter'
(grant numbers MR/S017216/1 \& MR/X006069/1). This project was also supported by the Science
and Technology Facilities Council [grant number ST/L00075X/1].
  This paper makes use of the following ALMA data:
  ADS/JAO.ALMA\#2016.1.00333.S, 2015.1.01136.S, and 2013.1.01241.S. ALMA is a partnership of ESO (representing
  its member states), NSF (USA) and NINS (Japan), together with NRC
  (Canada) and NSC and ASIAA (Taiwan) and KASI (Republic of Korea), in 
  cooperation with the Republic of Chile. The Joint ALMA Observatory is 
  operated by ESO, AUI/NRAO and NAOJ.
\end{acknowledgements}





\bibliographystyle{aa}
\bibliography{a383_oiii}

\end{document}